\documentclass[12pt]{article} %***
\usepackage[sectionbib]{natbib}
\usepackage{array,epsfig,fancyhdr,rotating}
\usepackage[]{hyperref}  %<----modified by Ivan
\usepackage{sectsty, secdot}
\sectionfont{\fontsize{12}{14pt plus.8pt minus .6pt}\selectfont}
\renewcommand{\theequation}{\thesection\arabic{equation}}
\subsectionfont{\fontsize{12}{14pt plus.8pt minus .6pt}\selectfont}
\usepackage{changepage}
\usepackage{booktabs}
\usepackage{algpseudocode}
\usepackage{amsmath}
\usepackage{amssymb}
\usepackage{amsfonts}
\usepackage{multirow}
\usepackage{amsthm}
\usepackage{setspace}
\usepackage{tikz}
\usepackage{subfigure}
\usetikzlibrary{shapes,decorations,arrows,calc,arrows.meta,fit,positioning}
\tikzset{
    -latex,auto,node distance =2 cm and 2 cm, on grid, very thick,
    state/.style ={ellipse, draw, minimum width = 1 cm},
    point/.style = {circle, draw, inner sep=0.04cm,fill,node contents={}},
    bidirected/.style={Latex-Latex,dashed},
    el/.style = {inner sep=2pt, align=left, sloped}
}

\newtheorem{theorem}{Theorem}

\newtheorem{corollary}{Corollary}

\theoremstyle{definition}

\newtheorem{example}{Example}

\newtheorem{condition}{Condition}
\newtheorem{assumption}{Assumption}
\newtheorem{step}{Step}
\usepackage{threeparttable}
\usepackage{color}
\usepackage{multirow}
\usepackage{xcolor}
\usepackage{bm}
\usepackage{enumitem}

\numberwithin{equation}{section}

\newcommand{\indep}{\perp \!\!\! \perp}

\def\T{{ \mathrm{\scriptscriptstyle T} }}

\allowdisplaybreaks
\begin{document}

%%%%%%%%%%%%%%%%%%%%%%%%%%%%%%%%%%%%%%%%%%%%%%%%%%%%%%%%%%%%%%%%%%%%%%%%%%%%%%%%%%%%%%%%%%%%%%%%%%%%%%%%%%%%%%%%%%%%%%%%%%%%
%%%%%%%%%%%%%%%%%%%%%%%%%%%%%%%%%%%%%%%%%%%%%%%%%%%%%%%%%%%%%%%%%%%%%%%%%%%%%%%%%%%%%%%%%%%%%%%%%%%%%%%%%%%%%%%%%%%%%%%%%%%%

\renewcommand{\baselinestretch}{2}

\markright{ \hbox{\footnotesize\rm Statistica Sinica
%{\footnotesize\bf 24} (201?), 000-000
}\hfill\\[-13pt]
\hbox{\footnotesize\rm
%\href{http://dx.doi.org/10.5705/ss.20??.???}{doi:http://dx.doi.org/10.5705/ss.20??.???}
}\hfill }

\markboth{\hfill{\footnotesize\rm FIRSTNAME1 LASTNAME1 AND FIRSTNAME2 LASTNAME2} \hfill}
{\hfill {\footnotesize\rm Auxiliary instrumental variables} \hfill}

\renewcommand{\thefootnote}{}
$\ $\par

%%%%%%%%%%%%%%%%%%%%%%%%%%%%%%%%%%%%%%%%%%%%%%%%%%%%%%%%%%%%%%%%%%%%%%%%%%%%%%%%%%%%%%%%%%%%%%%%%%%%%%%%%%%%%%%%%%%%%%%%%%%%

\fontsize{12}{14pt plus.8pt minus .6pt}\selectfont \vspace{0.8pc}
\centerline{\large\bf Identifying causal effects using instrumental variables}
\vspace{2pt}
\centerline{\large\bf from  the auxiliary dataset }
\vspace{.4cm}
\centerline{Kang Shuai$^{a}$, Shanshan Luo$^{b \ast}$\footnote{$\ast$Correspondence to: shanshanluo@btbu.edu.cn}, Wei Li$^c$ and Yangbo He$^{a }$}
\vspace{.4cm}
\centerline{$^a$School of Mathematical Sciences, Peking University}
\centerline{
$^b$School of Mathematics and Statistics, Beijing Technology and Business University}
\centerline{
$^c$Center for Applied Statistics and School of Statistics, Renmin University of China}
 \vspace{.55cm} \fontsize{9}{11.5pt plus.8pt minus.6pt}\selectfont

% long-term treatment effect
%%%%%%%%%%%%%%%%%%%%%%%%%%%%%%%%%%%%%%%%%%%%%%%%%%%%%%%%%%%%%%%%%%%%%%%%%%%%%%%%%%%%%%%%%%%%%%%%%%%%%%%%%%%%%%%%%%%%%%%%%%%%

\begin{quotation}
\noindent {\it Abstract: Instrumental variable approaches have gained popularity for estimating causal effects in the presence of unmeasured confounders. However, the availability of instrumental variables in the primary dataset is often challenged due to stringent and untestable assumptions.
This paper presents a novel method to identify and estimate causal effects by utilizing instrumental variables from the auxiliary dataset, incorporating a structural equation model, even in scenarios with nonlinear treatment effects. Our approach involves using two datasets: one called the primary dataset with joint observations of treatment and outcome, and another auxiliary dataset providing information about the instrument and treatment. Our strategy differs from most existing methods by not depending on the simultaneous measurements of  instrument and outcome. The central idea for identifying causal effects is to establish a valid substitute through the auxiliary dataset,  addressing unmeasured confounders.  This is achieved  by developing a control function and projecting it onto the function space spanned by the treatment variable. We then propose a three-step estimator for estimating   causal effects and derive its asymptotic results. We illustrate the proposed estimator through simulation studies, and the results demonstrate favorable performance. We also conduct  a real data analysis to evaluate the causal effect between vitamin D status and body mass index.}

\vspace{9pt}
    \par
\end{quotation}\par

\def\thefigure{\arabic{figure}}
\def\thetable{\arabic{table}}

\renewcommand{\theequation}{\thesection.\arabic{equation}}

\fontsize{12}{14pt plus.8pt minus .6pt}\selectfont

% =========================================================================
% Section 1

\section{Introduction}
\label{sec:intro}
Randomized controlled trials are generally considered as the gold standard for evaluating causal effects. However, conducting such trials may not always be feasible due to ethical concerns or practical constraints like cost.  In such situations, observational data can be used as an alternative for estimating causal effects.  The major challenge in observational studies is the presence of unmeasured confounders, which may often introduce bias and invalidate the  conclusions. Instrumental variables have been extensively employed to address such issues \citep{angrist1996identification, ogburn2015doubly}.
An instrumental variable is a pretreatment variable that satisfies certain criteria. It should be associated with the treatment variable, independent of unmeasured confounders, and only affect the outcome through the treatment variable. It is important to note that the instrumental variable can only be correlated with the treatment while not directly affecting the outcome.

Instrumental variable methods are commonly used to identify and estimate parameters in linear structural equation models (SEMs). When these models are correctly specified, the population average treatment effect   corresponds to a specific parameter in the SEMs and can be consistently estimated. The two-stage least squares (2SLS) method and the control function method are two prominent methods in the context of SEMs~\citep{goldberger1972structural,wooldridge2010econometric}. For both methods, the treatment is regressed on the instrument and baseline covariates in the first stage. However, in the second stage, the 2SLS method regresses the outcome on the predicted value of the treatment and baseline covariates, while the control function method regresses the outcome on the treatment, the residual from the first stage regression, and the baseline covariates. The 2SLS method aims to construct a function of the instrument and baseline covariates that is independent of unmeasured confounders, while the control function method incorporates the first stage residual, known as the control function, to control for unmeasured confounders \citep{imbens2007control,petrin2010control,wooldridge2015control}. Nonparametric or semiparametric identification of causal effects with an instrumental variable has also been sufficiently investigated in many researches, we refer readers to~\citet{angrist2013advances,ogburn2015doubly,wang2018bounded}.

In practice, due to various challenges, it is often difficult to gather complete information on treatments, outcomes, and instrumental variables.
 Consequently, the utilization of auxiliary data for identifying causal effects is a common strategy in instrumental variable analysis. One of the most widely used method  in this context is the two-sample instrumental variable framework
 \citep{angrist1992effect,arellano1992female}, extensively employed in econometrics, social sciences, and mendelian randomization studies \citep{inoue2010two,sun2022semiparametric,gamazon2015gene,zhao2019two,zhao2020statistical}.
The primary dataset provides  data   related to instruments and outcomes, whereas the auxiliary dataset provides information about the treatment and instrumental variables. The corresponding two-sample instrumental variable estimators utilize sample moments from both an instrument-treatment sample and an instrument-outcome sample to estimate causal effects. However, stringent assumptions might render a valid instrumental variable unavailable within the primary dataset. A more practical situation is that researchers can only jointly observe the treatment and the outcome that are subject to unmeasured  confounders.
For example, when assessing the effect of vitamin D deficiency on body mass index (BMI) in the primary dataset, commonly used instrumental variables like genetic factors might not be available. Meanwhile, there might be an auxiliary dataset that has collected data on both filaggrin mutation and vitamin D for other experimental purposes.  The filaggrin mutation plays a significant role in skin barrier function, which suggests a strong  association with vitamin D. Moreover, it is commonly acknowledged that the filaggrin mutation does not have a direct effect on BMI \citep{skaaby2013vitamin}. Consequently, the filaggrin mutation can be considered  as a candidate instrumental variable,  motivating further exploration of its potential to identify and estimate causal effect  within the primary dataset.

 In this paper, we introduce a novel approach for identifying and estimating treatment effects with an instrumental variable from the auxiliary dataset. Unlike conventional two-sample instrumental variable methods, we do not require the simultaneous measurements of  instrument and outcome in the primary dataset.
 Instead, for a specific treatment variable that may be confounded in the primary dataset, our approach relies on the presence of an auxiliary dataset that includes a valid instrument and the treatment variable.
The presence of unobserved confounders and the absence of instrumental variables commonly occur in the primary dataset, motivating us to utilize the  instrumental variables in the auxiliary dataset for identifying causal effects.  This supplementary dataset is typically available from various data sources, containing randomized experiments or observational studies.
%In the instrumental variable analysis, the treatment variable typically captures a short-term impact, while the outcome variable reflects a longer-term impact.
%Researchers generally find it more feasible to collect data on the short-term impact (i.e., the treatment variable) compared to the long-term impact (i.e., the outcome variable).}
Although such ideas are straightforward and practical, but as far as we know it has not appeared in the literature. Even within the classical SEM framework, how can we effectively evaluate treatment effects without simultaneous measurements of both   instrument and outcome?

The control function approach within the SEMs  provides valuable insights into the effective utilization of instrumental variables in the auxiliary dataset.
%auxiliary data that solely   includes both the instrument and treatment.
In this paper, we adopt the control function perspective and establish a set of sufficient conditions to guarantee the identification of treatment effects. Specifically, by projecting the control function onto the function space spanned by the treatment variable, we can construct a potentially valid substitute for   unmeasured confounders, namely, the control function projection. By incorporating the control function projection variable into the outcome model, we can effectively  remove the impact of unmeasured confounders, leading to the  identification of the treatment effect. Importantly, our method is also applicable to nonlinear treatment effects, and the conditions considered in this paper are expected to be no stronger than the similar assumptions made by  ~\citet{imbens2007control} and  \citet{guo2016control}. Based on the identification results, we propose a three-step estimator for estimating the  causal  effect. We demonstrate its consistency of rate $n^{-1/2}$ under certain regularity conditions, subject to the requirement that the control function projection exhibits a uniform convergence rate of at least $n^{-1/4}$.

The remaining sections of this paper are organized as follows. In Section~\ref{model}, we present the notation and outline the proposed model.  Section~\ref{ident} presents a brief review of the control function and  introduces  sufficient conditions for identifying causal effects.  Section~\ref{est} provides a three-step estimator and establishes its asymptotic results. To evaluate the empirical performance of the proposed estimator, we conduct a simulation study in Section~\ref{simu}. Furthermore, in Section~\ref{app}, we apply the proposed estimation procedure to real-world vitamin D datasets. The extension and further discussions are presented in Section~\ref{relax} and~\ref{dis}, respectively.

\section{Notation  and Model}\label{model}
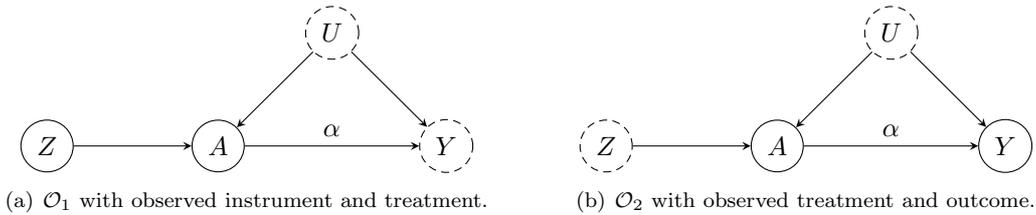
\begin{figure}[t]
    \centering
    \subfigure[$\mathcal{O}_1$ with observed instrument and treatment.]{\begin{minipage}{6cm}
    \centering
\begin{tikzpicture}[>=stealth, scale=1.5]
    \node[draw, circle] (Z) at (0.5, 1) {$Z$};
    \node[draw, circle] (A) at (2, 1) {$A$};
    \node[draw, circle, densely dashed] (Y) at (4, 1) {$Y$};
    \node[draw, circle, densely dashed] (U) at (3, 2) {$U$};
    \draw[->] (Z) -- (A);
    \draw[->] (A) -- node[above] {$\boldsymbol \alpha$} (Y);
    \draw[->] (U) -- (A);
    \draw[->] (U) -- (Y);
\end{tikzpicture}\vspace{2mm}
    \end{minipage}}
    \subfigure[$\mathcal{O}_2$ with observed treatment and outcome.]{
    \begin{minipage}{6cm}
    \centering
\begin{tikzpicture}[>=stealth, scale=1.5]
    \node[draw, circle, densely dashed] (Z) at (0.5, 1) {$Z$};
    \node[draw, circle] (A) at (2, 1) {$A$};
    \node[draw, circle] (Y) at (4, 1) {$Y$};
    \node[draw, circle, densely dashed] (U) at (3, 2) {$U$};
    \draw[->] (Z) -- (A);
    \draw[->] (A) -- node[above] {$\boldsymbol \alpha$} (Y);
    \draw[->] (U) -- (A);
    \draw[->] (U) -- (Y);
\end{tikzpicture}\vspace{2mm}
    \end{minipage}
    }
    \caption{Causal diagrams for two different datasets, one without outcome and the other  without instrument. The dashed circles indicate that the corresponding variable is unobserved in the dataset. For simplicity, we have omitted the baseline covariates and selection indicator.
    }
    \label{fig:data}
\end{figure}

%Let $A$ represent a continuous treatment, $Y$ a continuous outcome, $U$ a vector of unmeasured confounders, and $Z$ a discrete or continuous instrumental variable.
We assume that $A$ denotes a scalar continuous treatment, $Y$ denotes a scalar continuous outcome, $ \boldsymbol U$ denotes a $t$-dimensional vector of unmeasured confounders, and $Z$ denotes an instrumental variable that can be either discrete or continuous.
For notational simplicity, we condition on covariates implicitly and firstly omit them below. We adopt the potential outcomes framework to define causal effects and make the stable unit treatment value assumption (SUTVA) throughout the paper, that is, there is only one version of the potential outcomes and there is no interference between units \citep{rubin1980randomization}. The SUTVA allows us to uniquely define the potential outcome $Y_a$ for the outcome   if the treatment  $A$ is set to be $a$. Suppose we have two datasets: the primary dataset, denoted as $\mathcal{O}_2 = \{ (A_j, Y_j) : j \in \mathcal{S}_2 \}$ with $n_2 = \lvert \mathcal{S}_2 \rvert$ samples, and the auxiliary dataset, denoted as $\mathcal{O}_1 = \{(Z_i, A_i) : i \in \mathcal{S}_1 \}$ with $n_1 = \lvert \mathcal{S}_1 \rvert$ samples. For simplicity, we assume that $\mathcal{S}_1 \cap \mathcal{S}_2 = \emptyset$, and we denote $r = \lim_{n_2 \to \infty} {n_2}/{n_1} \in (0, \infty)$. We assume that all data are independently and identically distributed for $i \in \mathcal{S} = \mathcal{S}_1 \cup \mathcal{S}_2$. Let $R_i$ be an indicator variable, where $R_i=1$ if the $i$th unit is from the primary dataset and $R_i=0$ otherwise. In our analysis, we initially assume that the selection indicator $R_i$ is independent of the variables $(Z_i, A_i, Y_i)$, and later explore the potential relaxation of this assumption in Section \ref{relax}. Therefore, the combined set of observed data can be represented as $\mathcal{O} = \{( R_i, R_i Y_i,   Z_i-Z_i R_i, A_i ): i \in \mathcal{S}\}$, consisting of total $n = n_1 + n_2$ samples.
We propose the following model, where $g(\cdot)$ represents a vector of known linearly independent functions:
\begin{equation}
    \label{eqn:modelforAY}
    \begin{aligned}
Y &= \boldsymbol \alpha^{\T} \boldsymbol g(A) + \boldsymbol \beta^{\T} 
\boldsymbol U + \eta,  \\
A &= m(Z) + \boldsymbol l^{\T} \boldsymbol U + \varepsilon,
\end{aligned}
\end{equation}
where two error terms and unmeasured confounders satisfying $\varepsilon \indep (Z, \boldsymbol U)$ and $\eta \indep (Z, A, \boldsymbol U)$, $\mathrm{var}( \boldsymbol U) = \boldsymbol I_t$ and $\mathrm{var}(\varepsilon) = \sigma^2$. Here, $\eta \indep (Z,A, \boldsymbol U)$ can be relaxed to $E(\eta\mid Z,A, \boldsymbol U)=0$. However, the independence of $\varepsilon$ and $(Z, \boldsymbol U)$ is important to guarantee identification of $\boldsymbol \alpha$ as illustrated by several examples in Section~\ref{ssec:identification}. Without loss of generality, throughout this paper,  we assume that $\{ \boldsymbol U, \eta, \varepsilon, A, \boldsymbol g(A), Y \}$ have zero means. Model \eqref{eqn:modelforAY} includes a class of structural equation models (SEMs). For instance, when $\boldsymbol g(A) = A$ and $ \alpha \in \mathbb{R}^1$, model \eqref{eqn:modelforAY} is the widely used linear structural equation model  in instrumental variable analysis, where the parameter  $\alpha$ represents the causal effect of $A$ on $Y$ for a unit increase of $A$.  Model \eqref{eqn:modelforAY} also includes scenarios where the function $\boldsymbol g(A)$ may contain nonlinear terms of $A$, and such nonlinear treatment effects are frequently observed in practice, as discussed by~\citet{guo2016control} and \citet{li2020causal}.
When $Z$ is binary, the term $m(Z)$ can be characterized by the saturated model, $m(Z)=\gamma_0 + \gamma_1 Z$, and it must be linear with respect to $Z$. Figure~\ref{fig:data} provides a graphical illustration using two observational datasets $\mathcal{O}_1$ and $\mathcal{O}_2$.

Given model \eqref{eqn:modelforAY}, the causal effect of treatment $A$ on outcome $Y$ can be quantified by comparing the potential outcomes when we change $A$ from $a$ to $a'$, that is,
$
Y_{a'}- Y_{a} = \boldsymbol \alpha^{\T} \{ \boldsymbol g(a') - \boldsymbol g(a) \}.
$
Throughout this paper, our primary focus is on identifying and estimating the parameter $\boldsymbol \alpha$. As mentioned earlier, in common linear models, $\boldsymbol \alpha$ can represent the true causal effect. However, in general nonlinear models, the parameter $\boldsymbol \alpha$ represents the change in the potential outcome corresponding to the treatment changing from $a$ to $a'$.
When $\boldsymbol g(A)$ is a vector, even with the complete information of $(Z,A,Y)$ available, the two-stage least squares   method may not yield a consistent estimator of $\boldsymbol \alpha$ without additional conditions~\citep{guo2016control}.
Specifically, within the two-stage least squares  framework, it is crucial to ensure that the conditional expectation vector $E\{ \boldsymbol g(A)\mid Z \}$ is linearly independent in order to estimate the parameter $\boldsymbol \alpha$.  However, this condition can be easily violated when the instrument $Z$ is binary, making direct regression of $Y$ on $E\{ \boldsymbol g(A)\mid Z \}$
 infeasible. In contrast, the control function approach provides an alternative method for estimating $\boldsymbol \alpha$ even with a binary instrument $Z$. By incorporating the vector $\boldsymbol g(A)$ into $
 \{ \boldsymbol g(A)^{\T}, A - m(Z) \}^{\T}$, we ensure that the resulting vector remains linearly independent, which further leads to more precise estimates of causal parameter $\boldsymbol \alpha$~\citep{imbens2007control}.

\section{Identification}\label{ident}
\subsection{Review of control function approach}\label{rev}
In this section, we will give a brief review of the control function approach, which provides a different perspective for estimating possibly nonlinear treatment effects compared to the  two-stage least squares method. The control function approach is a widely adopted method for identifying and estimating  linear or nonlinear treatment effects with instrumental variables~\citep{heckman1976common,rivers1988limited, imbens2007control, petrin2011revisiting, cai2011two, wooldridge2015control, guo2016control, li2020causal}.  
 In this section, we start by considering a straightforward scenario involving a complete dataset, where measurements for the instrumental variable, treatment variable,  and outcome variable are all collected simultaneously. We will begin by reviewing the control function approach within the context of a linear outcome model, and then proceed to extend it to the nonlinear case. Specifically, in the simplest case where $\boldsymbol g(A)=A$, we define $\boldsymbol U_{\mathrm{proj}}$ to signify the linear projection of $\boldsymbol U$ on $A-m(Z)$:
 \begin{align*}
\boldsymbol U_{\mathrm{proj}} 
&= \mathrm{cov}\{ \boldsymbol U,A-m(Z) \} \cdot \mathrm{cov}^{-1} \{A-m(Z) \} \cdot \{ A - m(Z) \} \\
&= \frac{ \boldsymbol l}{ \boldsymbol l^{\T} \boldsymbol l + \sigma^2} \{ A - m(Z) \}.
\end{align*}
The outcome model can be rewritten as follows:
\begin{equation}
    \label{eq:outcome2}
\begin{aligned}
Y 
&= \alpha A + \boldsymbol \beta^{\T} \boldsymbol U_{\mathrm{proj}} + \boldsymbol \beta^{\T} ( \boldsymbol U - \boldsymbol U_{\mathrm{proj}}) + \eta  \\
&= \alpha A + \frac{ \boldsymbol \beta^{\T} \boldsymbol l}{ \boldsymbol l^{\T} \boldsymbol l + \sigma^2} \{ A - m(Z) \} + \boldsymbol \beta^{\T} ( \boldsymbol U - \boldsymbol U_{\mathrm{proj}}) + \eta.
\end{aligned}
\end{equation}
By the definition of linear projection and $Z \indep ( \boldsymbol U,\varepsilon)$, we  have that  $\mathrm{cov} \{ A-m(Z) ,\boldsymbol U- \boldsymbol U_{\mathrm{proj}}\} = 0$ and
$$
\mathrm{cov} \left\{ m(Z), \boldsymbol U- \boldsymbol U_{\mathrm{proj}} \right\} = \mathrm{cov} \left\{ m(Z), \boldsymbol U- \frac{ \boldsymbol l}{ \boldsymbol l^{\T} \boldsymbol l + \sigma^2}( \boldsymbol l^{\T} \boldsymbol U + \varepsilon) \right\}=0.
$$
The above two conditions also imply the zero covariance between $A$ and $\boldsymbol U - \boldsymbol U_{\mathrm{proj}}$, that is, $\mathrm{cov} \{ A , \boldsymbol U - \boldsymbol U_{\mathrm{proj}}\} = 0$. Therefore,  $\boldsymbol \beta^{\T} ( \boldsymbol U- \boldsymbol U_{\mathrm{proj}}) + \eta$  in \eqref{eq:outcome2} can be regarded as a new error term, which is uncorrelated with $A$ and $A - m(Z)$. Hence, if we regress $Y$ on $A$ and $A-m(Z)$, then the coefficient of $A$ will be equal to the true causal effect $ \alpha$. This method is the so-called control function approach, which has been previously discussed in~\citet{heckman1976common}.

However, the outcome model may also be nonlinear with respect to the treatment, the direct regression procedure may not be sufficient for identifying causal parameter $\boldsymbol \alpha$
due to the potential non-zero covariance between $\boldsymbol g(A)$ and $\boldsymbol U - \boldsymbol U_{\mathrm{proj}}$.  In such cases, the identifiability condition outlined by~\citet{imbens2007control}, and further discussed by~\citet{guo2016control} and~\citet{li2020causal}, becomes essential for identifying and estimating the parameter $\boldsymbol \alpha$. We summarize the condition as follows:
\begin{condition}\label{C1}
%The unobserved part $ \beta^{\T} U + \eta$ in the outcome model and $l^{\T} U +\varepsilon$ in the treatment model satisfy:
$
E( \boldsymbol \beta^\T \boldsymbol U + \eta \mid \boldsymbol l^\T \boldsymbol  U + \varepsilon ) = \rho ( \boldsymbol l^\T \boldsymbol U + \varepsilon),
%E( \beta^{\T} U + \eta \mid l^{\T} U + \varepsilon ) = \rho (l^{\T} U + \varepsilon)
$
where $\rho$ is some constant.
\end{condition}
A sufficient condition for Condition \ref{C1} is that the conditional expectation $E( \boldsymbol U \mid \boldsymbol l^{\T} \boldsymbol U + \varepsilon)$ is linear in $\boldsymbol l^{\T} \boldsymbol U + \varepsilon$. Condition \ref{C1} implies that regressing $Y$ on $\boldsymbol g(A)$ and the residual $A - m(Z)$ will give the consistent estimator of $\boldsymbol \alpha$, as shown below:
\begin{equation}
    \label{eq:nonlinear-CF}
    \begin{aligned}
    E(Y\mid Z,A) &= \boldsymbol \alpha^{\T} \boldsymbol g(A) + E( \boldsymbol \beta^{\T} \boldsymbol U + \eta \mid Z,A ) \\&= \boldsymbol \alpha^{\T} \boldsymbol g(A) + E( \boldsymbol \beta^\T \boldsymbol U + \eta \mid \boldsymbol l^\T \boldsymbol  U + \varepsilon )\\& = \boldsymbol \alpha^{\T} \boldsymbol g(A) + \rho \{ A -m(Z) \},
\end{aligned}
\end{equation}
 where the last equality  holds due to model~\eqref{eqn:modelforAY}.
 However, equation \eqref{eq:nonlinear-CF} still requires the joint measurements of the instrument and   outcome, and it does not directly address the specific problem encountered in Figure \ref{fig:data}.
In the upcoming section, we present a series of novel conditions to identify the causal parameter $\boldsymbol \alpha$ using the combined dataset $\mathcal{O}$ without directly imposing Condition~\ref{C1}. %, {\red which are weaker than requiring $E(U \mid l^{\T} U + \varepsilon)$ to be linear in $l^{\T} U + \varepsilon$}.
\subsection{Identification using auxiliary dataset}
\label{ssec:identification}
To identify the causal effect $\boldsymbol \alpha$, we introduce a conditional expectation $C(A)$, defined as $C(A) = E\{ A - m(Z) \mid A \} = E( \boldsymbol l^{\T} \boldsymbol U + \varepsilon \mid A)$. Obviously,  the conditional expectation $C(A) = A - E\{ E(A\mid Z,R=0) \mid A,R=0\}$ can be identified using the auxiliary dataset $\mathcal{O}_1$. The term $A - m(Z) = \boldsymbol l^{\T} \boldsymbol U + \varepsilon$ is known as the control function \citep{heckman1976common}, and $C(A)$ represents the projection of the control function onto the function space spanned by the treatment variable $A$. Intuitively, the control function projection variable $C(A)$ captures the information of unmeasured confounders $\boldsymbol U$ entailed by treatment $A$, and can be considered as a potentially valid substitute for unmeasured confounders $\boldsymbol U$. We consider the following regular condition:

%Let $C(A) = E\{ A - m(Z) \mid A \} = E( l^{\T} U + \varepsilon \mid A)$, and we consider the regular condition for identifying $\boldsymbol \alpha$:
\begin{assumption}\label{A1}
The vector $\boldsymbol g(A)$ and $C(A)$ are linearly independent.
\end{assumption}
The plausibility of Assumption~\ref{A1} can be assessed using the observed data, since the conditional expectation $C(A)  $ is identifiable  and the vector $\boldsymbol g(A)$ is  known. When $\boldsymbol g(A)=A$, this assumption implies that the term $E\{ m(Z)\mid A \}$ is nonlinear with respect to $A$, which is often easily satisfied when $Z$ is binary.

As shown in \eqref{eq:nonlinear-CF}, the control function approach for a linear treatment effect with the full dataset $\mathcal{F} = \{ (Z_i,A_i,Y_i): i=1,...,n \}$ does not require any additional assumptions. However, when $\boldsymbol g(A)$ contains a nonlinear term of $A$, even with the full dataset $\mathcal{F}$, additional assumptions must be introduced, such as Condition~\ref{C1} \citep{guo2016control}.  Nevertheless,   Condition~\ref{C1} does not address the challenge of lacking the joint observation of instrumental and outcome variables. To address this issue, we propose the following assumption:
\begin{assumption}\label{A2}
$E(\boldsymbol U \mid A) = \boldsymbol v E(\varepsilon \mid A$), where $\boldsymbol v$ is a constant vector.
\end{assumption}
It is important to mention that the constant vector  $\boldsymbol v$ can be calculated as $\boldsymbol v = \boldsymbol l/\sigma^2$, based on the fact $E(A \boldsymbol U) = \boldsymbol v E( A \varepsilon)$.
Assumption~\ref{A2} plays a pivotal role in our subsequent discussions on identifiability. This assumption is crucial not only for ensuring the identification of the causal parameter $\boldsymbol \alpha$, but also for addressing the data fusion problem in cases where the instrument and outcome variables are not observed jointly.
%Assumption~\ref{A2} not only ensures the identification of the causal parameter $\boldsymbol \alpha$ but also allows us to address the data fusion problem when the instrument and outcome variables are not jointly observed, which is pivotal in our subsequent discussions on identifiability.
In the context of model \eqref{eqn:modelforAY} with a scalar unmeasured confounder, Assumption~\ref{A2} is indeed weaker compared to Condition~\ref{C1}, as illustrated in the following Example \ref{exm_1}.
We also provide a mathematical relaxation of this assumption in the supplementary material. Additionally, Assumption \ref{A2} imposes constraints on the joint distribution of the unmeasured confounder $\boldsymbol U$ and the noise $\varepsilon$, requiring the two corresponding conditional expectations with respect to $A$ to be proportional.  A similar idea has also been previously considered by~\citet{shuai2024identification} for identifying causal effects in the presence of unmeasured confounders. It is important to note that the conditional expectation $E(\boldsymbol U\mid A) = E_{Z\mid A}\{  E(\boldsymbol U\mid Z, \boldsymbol l^\T \boldsymbol U + \varepsilon)  \} = E_{Z\mid A}\{  E(\boldsymbol U\mid \boldsymbol l^\T \boldsymbol U + \varepsilon)  \} $ and $E(\varepsilon\mid A) = E_{Z\mid A}\{ E(\varepsilon \mid \boldsymbol l^\T \boldsymbol U + \varepsilon) \}$, which implies that we only need to derive the proportional equality as a sufficient condition when conditional on $\boldsymbol l^\T \boldsymbol U + \varepsilon$. When this holds, $E(\boldsymbol U \mid A)$ will be proportional to $E(\varepsilon \mid A)$. We now provide additional examples for illustrating the plausibility and applicability of Assumption~\ref{A2}.

\begin{example} \label{exm_1}
Under our setting, Assumption 4 in~\citet{guo2016control} can be interpreted as:
$$
E(\beta U + \eta  \mid l U + \varepsilon) = \rho (l U + \varepsilon),
$$
where $\rho \in \mathbb{R}^1 \backslash \{ 0 \}$ is a constant. This equality is essentially Condition~\ref{C1} stated in Section~\ref{rev}.
This entails that $E( U \mid l U + \varepsilon ) $ is linear in $l U + \varepsilon$. Since $E( l U + \varepsilon \mid lU + \varepsilon)$ is obviously linear in $lU + \varepsilon$, we thus have $E( \varepsilon \mid l U + \varepsilon ) $ is also linear in $l U + \varepsilon$. Given $Z \indep (U,\varepsilon)$, we immediately have:
\begin{align*}
     E\{ U\mid  m(Z),A  \} &= E\{ U\mid  m(Z), l U + \varepsilon \} = E( U\mid l U + \varepsilon ),\\
     E\{ \varepsilon\mid m(Z), A \} &= E\{ \varepsilon \mid  m(Z), l U + \varepsilon \} = E( \varepsilon \mid l U + \varepsilon ).
\end{align*}
 Therefore, two conditional expectations $E\{ U\mid m(Z), A \}$ and $E\{ \varepsilon\mid m(Z),A \}$ should also be linear in $l U + \varepsilon$. Then there must exist some constant $v$ satisfying
 $$
 E\{ U\mid m(Z), A \} = v E\{ \varepsilon\mid m(Z), A \}.
 $$
  This implies Assumption~\ref{A2}.
\end{example}
Example \ref{exm_1} illustrates that the conditions considered in this paper are expected to be no stricter than Condition \ref{C1} within the context of a scalar unmeasured confounder  \citep{imbens2007control,guo2016control}.  We next provide specific examples of the possible joint distribution    of $(\boldsymbol U^{\T},\varepsilon)^{\T}$ based on Assumption~\ref{A2} in a more general case.

\begin{example}[Elliptical Distribution] \label{exm_2}
Suppose the joint probability density function of $ \boldsymbol \zeta = (\boldsymbol U^{\T},\varepsilon)^{\T}$ has the following elliptical distributional form:
$$
p_{\boldsymbol \zeta}(x) = k \cdot \varphi \{ ( \boldsymbol x - \boldsymbol \mu)^{\T} \boldsymbol \Sigma^{-1} ( \boldsymbol x - \boldsymbol \mu) \},
$$
where $k$ is the normalizing constant, $\boldsymbol \mu$ is the mean vector, $\boldsymbol \Sigma$ is the covariance matrix, and $\varphi (\cdot)$ is some known function. By  Theorem 2.18 in~\citet{fang2018symmetric}, we have:
$$
E(\boldsymbol \zeta \mid \boldsymbol B^{\T} \boldsymbol \zeta ) = \boldsymbol a_{\boldsymbol B} + \boldsymbol M_{\boldsymbol B} \boldsymbol B^{\T} \zeta,
$$
where $\boldsymbol a_{\boldsymbol B} \in \mathbb{R}^{t+1}$ is some constant vector and $ \boldsymbol B \in \mathbb{R}^{(t+1) \times r}$ is some constant matrix with full column rank. Because $E(\boldsymbol \zeta) = \boldsymbol 0$, we must have $\boldsymbol a_{\boldsymbol B} = \boldsymbol 0$. Specifically, let $\boldsymbol B= (\boldsymbol l^{\T}, 1)^{\T}  $, this implies:
$$
E(\boldsymbol \zeta \mid \boldsymbol l^{\T} \boldsymbol U + \varepsilon) =\boldsymbol  M_{\boldsymbol B} (\boldsymbol l^{\T} \boldsymbol U + \varepsilon).
$$
Thus, there must exist some constant vector $\boldsymbol v \in \mathbb{R}^t$ satisfying
$$
%%%%%%%% FROM NOW ON
E\{ \boldsymbol U\mid m(Z), A \} = E(\boldsymbol U\mid \boldsymbol l^{\T} \boldsymbol U + \varepsilon) = v E(\varepsilon\mid \boldsymbol l^{\T} \boldsymbol U + \varepsilon) = v E\{ \varepsilon \mid m(Z), A \}.
$$
 This implies Assumption~\ref{A2}.
\end{example}

Elliptical distributions play an important role in various fields, including portfolio theory~\citep{owen1983class}. Example~\ref{exm_2} covers a broad class of distributions satisfying Assumption~\ref{A2}, such as the multivariate normal distribution, multivariate $t$-distribution, multivariate Laplace distribution, and so on~\citep{fang2018symmetric}. The elliptical distribution can be either bounded or unbounded, depending on the choice of the function $\varphi(\cdot)$. Additionally, the Pearson system of distributions can also fulfill Assumption~\ref{A2}~\citep{kotz2004continuous}. The following example provides another case that guarantees Assumption \ref{A2} without requiring a specific distributional form.

\begin{example} \label{exm_3}
Let $\boldsymbol l = (1,...,1)^{\T}$,   Assumption~\ref{A2} can also be satisfied when $U_1,...,U_t,\varepsilon $ are independent and identically  distributed. Specifically, we have
$$
E(U_1\mid \boldsymbol l^{\T} \boldsymbol U + \varepsilon) = \cdot\cdot\cdot = E(U_t\mid \boldsymbol l^{\T} \boldsymbol U + \varepsilon) = E(\varepsilon\mid  \boldsymbol l^{\T} \boldsymbol U + \varepsilon),
$$
where $\boldsymbol U = (U_1,...,U_t)^{\T}$. This means
$$
E( \boldsymbol U \mid \boldsymbol l^{\T} \boldsymbol U + \varepsilon ) = \boldsymbol v E( \varepsilon \mid \boldsymbol l^{\T} \boldsymbol U + \varepsilon) \text{ and } \boldsymbol v = \boldsymbol l/ \sigma^2.
$$
This implies Assumption~\ref{A2}.
\end{example}

The example above illustrates that if the underlying causal mechanisms of unmeasured variables $U_1, U_2, ..., U_t$, and $\varepsilon$ are identical with respect to the treatment variable $A$, then Assumption~\ref{A2} also hold.  As a result, the causal parameter $\boldsymbol \alpha$ can be identified without imposing any  restrictions on their joint distribution.

\begin{theorem}\label{thm1}
Under Assumptions~\ref{A1}-\ref{A2}, the causal parameter $\boldsymbol \alpha$ is  identified by
\begin{align}
    \label{eq:iden-alpha}
    \boldsymbol \alpha = \boldsymbol D_{11} E\{ \boldsymbol g(A) Y \mid R=1 \} + \boldsymbol D_{12} E\{ C(A) Y \mid R=1 \},
\end{align}
where   $\boldsymbol D_{11}$ and $\boldsymbol D_{12}$ are the corresponding block matrices of the variance matrix $\boldsymbol D$:
$$
\boldsymbol D =
\left[ E\left\{ \boldsymbol h(A) \boldsymbol h(A)^{\T} \right\} \right]^{-1}
=
\begin{pmatrix}
\boldsymbol D_{11}  & \boldsymbol D_{12} \\
 \boldsymbol D_{21}  &  \boldsymbol D_{22}
\end{pmatrix}
, \quad \boldsymbol h(A) = \{ \boldsymbol g(A)^{\T}, C(A) \}^{\T}. $$
\end{theorem}
As discussed below Assumption \ref{A1}, the control function projection $C(A)$ can be obtained using the auxiliary dataset $\mathcal{O}_1$. Once we have derived the projection $C(A)$, we can directly perform a regression of $Y$ on $h(A)$ in the dataset $\mathcal{O}_2$. The coefficient of $\boldsymbol g(A)$ can then be used to identify the causal parameter $\boldsymbol \alpha$. The theorem essentially requires that either the instrument or the outcome variable is missing completely at random (MCAR), namely, $R\indep (Z,A,Y)$. However, our situation is more complex compared to the traditional MCAR framework. Whether restricted to the subpopulation with $R=1$ or $R=0$, we cannot obtain a complete observed dataset, making it challenging to directly apply standard missing data analysis methods \citep{rubin1976inference}. 
To partially relax this  MCAR assumption, we will address the same identification issue even when the selection indicator $R$ depends on certain observed variables, as discussed in Section \ref{relax}. By the way, we provide the following Corollary to include baseline covariates $W$ into analysis.

\begin{corollary}
\label{coro:cov}
    Let $W$ represent the baseline covariates, replace Assumption~\ref{A2} by $E(U\mid A,W) = v E(\varepsilon\mid A,W)$ and assume $g(A),E\{ m(Z,W) \mid A, W \}$ are linearly independent and the following model holds
   \begin{equation}
    \label{eqn:newmodelforAY}
    \begin{aligned}
Y &= \boldsymbol \alpha^{\T} \boldsymbol g(A) + \boldsymbol \beta^{\T} 
\boldsymbol U + h(W) + \eta,  \\
A &= m(Z,W) + \boldsymbol l^{\T} \boldsymbol U + \varepsilon,
\end{aligned}
\end{equation}
where $\boldsymbol g(\cdot)$ is known, $h(\cdot)$ is unknown and $(\boldsymbol U, \varepsilon) \indep (Z,W)$. Then we can conclude that $\boldsymbol \alpha$ is identifiable.
\end{corollary}

\section{Estimation and Inference}\label{est}
% {\red
%  From Theorem~\ref{thm1}, the identification formula can be constructed using the combined dataset $\mathcal{O}$.
% The auxiliary dataset $\mathcal{O}_1$ is mainly leveraged for estimating the control function projection variable $C(A)$.}
% \subsection{General estimation strategies and theoretical properties}
In this section, we omit the covariates and first describe an estimation procedure in full generality.
\begin{step}
 Obtain a  treatment model $\hat{m}(Z)$ for $m(Z)$ based on the auxiliary dataset $\mathcal{O}_1$.
\end{step}

\begin{step} Project the residue $A-\hat{m}(Z)$   onto the function space of $A$ to derive the control function projection estimator $\hat{C}(A)$ for $C(A)$ based on the auxiliary dataset $\mathcal{O}_1$.
\end{step}

\begin{step} Plug the estimate   $\hat{C}(A)$  into \eqref{eq:iden-alpha}  to estimate $\boldsymbol \alpha$ based on the primary dataset $\mathcal{O}_2$.
\end{step}
 In Step 1, estimating the treatment model $ {m}(Z)=E(A\mid Z)$ can be achieved parametrically or nonparametrically using standard density estimation techniques. This step is relatively straightforward. However, Step 2 poses a more significant challenge as it involves solving a reverse estimation problem for $C(A)=A-E\{m(Z)\mid A,R=0\}$. In most literature, the primary focus remains on estimating $m(Z)$, representing the forward estimation problem in Step 1, which also aligns with the data generation mechanism. Nevertheless, even if a parametric model for $m(Z)$ is known, or in the case where $Z$ is binary, obtaining the true parametric model for $C(A)$ can still be challenging. In practice, we recommend utilizing nonparametric estimation techniques.
 %At the end of this  subsection, we will describe some nonparametric estimation techniques applicable to the linear model setting of $m(Z)$. These techniques shed light on solving the most classical SEM. Without the linear structure of $m(Z)$, we refer reders to the two-step sieve  M  estimator previously discussed by~\citet{hahn2018nonparametric}.
 Estimating Step 3 is standard in causal inference problems. We will outline some regularity conditions required for Step 2, which are sufficient to ensure the asymptotic results for the causal  parameter $\boldsymbol \alpha$.

 \begin{assumption}
\label{assump:regular1}  (i) $\hat{C} ( A)$ is uniformly consistent, namely, $
\sup_{A\in \mathcal{A}} \rvert \hat{C} (A) - C(A) \rvert = o_p(1)$; and (ii)   $\sup_{A\in \mathcal{A}} \lvert  \hat{C} ( A)  -{C} ( A)  \rvert = o_p ( n_1^{-1/4} )$.
\end{assumption}
The  conditions imposed in
Assumption \ref{assump:regular1}  are standard regularity conditions used to establish the asymptotic results.
Assumption \ref{assump:regular1}(ii)  is commonly imposed in the causal inference literature to derive the asymptotic distribution of the  estimators when the nuisance functions are estimated with certain convergence rates.  For example, if $\hat{C}( A)$ is estimated based on a correctly specified parametric model,  $\hat{C}( A)$  is $\sqrt{n_1}$-consistent by using maximum likelihood estimation or generalized moments of estimation~\citep{hansen1982large}. When $m(Z)$ and $\hat{C}( A)$
 are estimated nonparametrically, it is often expected that the estimators can achieve the rate $o_p(n_1^{-1/4})$. Specifically, we can adopt the classical linear sieve methods with certain linearly independent basis functions. Except for the sieve method, popular machine learning approaches, including kernel regression or random forest, can also be easily implemented to achieve the desired square root consistency. Furthermore, while sample splitting or cross-fitting techniques are not required for our estimation, they provide a valuable perspective for alleviating some technical conditions in future work  \citep{chernozhukov2018double}. %%这个地方我修改了一些内容
 Although Assumption \ref{assump:regular1}(ii)  related to the sample size $n_1$, it is important to note that we assume the ratio $n_2/n_1$ of the two sample sizes tends to a constant $r$. This implies that condition  $\sup_{A\in \mathcal{A}} \lvert  \hat{C} ( A)  -{C} ( A)  \rvert = o_p ( n_2^{-1/4} )$ also holds.

In the final part of this section,   we will outline  simpler conditions to ensure that Assumption \ref{assump:regular1} is satisfied within the context of a linear model framework for $m(Z)$.   These conditions provide insights into  solving the classical linear structural equation model more effectively. Without the linear structure of $m(Z)$, we refer readers to the two-step sieve  M  estimator previously discussed by~\citet{hahn2018nonparametric}.

\begin{theorem}\label{thm2}
Under Assumptions \ref{assump:regular1}(i), we have that $\hat{\boldsymbol \alpha}$ is consistent for $\boldsymbol \alpha$ as $n_2\to \infty$. Additionally, suppose Assumption \ref{assump:regular1}(ii) holds, then $\hat{\boldsymbol \alpha}$ is asymptotically normal,  namely,
 $
\sqrt{n_2} ( \hat{\boldsymbol \alpha} - \boldsymbol \alpha ) \overset{d}{\to} N(0,\boldsymbol V)
$ as $n_2\to \infty$,
where
\begin{gather*}
   \boldsymbol V =  \mathrm{ var} \bigg\{ \boldsymbol D_{11} \boldsymbol g(A)Y + \boldsymbol D_{12} C(A)Y + {  \frac{\partial \boldsymbol \alpha(\boldsymbol \mu)}{ \partial \boldsymbol \mu}} X  \bigg\},\\
\boldsymbol \alpha( \boldsymbol \mu) = \boldsymbol D_{11}  E\{ \boldsymbol g(A) Y \} + \boldsymbol D_{12}   E\{ C(A) Y \}, \\
\boldsymbol X = \big[ \mathrm{ vec} \{ \boldsymbol g(A)\boldsymbol g(A)^{\T} \}^{\T}, \boldsymbol g(A)^{\T} C(A),C(A)^2 \big]^{\T}, ~~~ \boldsymbol \mu = E(\boldsymbol X).
\end{gather*}
\end{theorem}
The above theorem demonstrates that the consistency of $\hat{\boldsymbol \alpha}$ can be achieved when a uniformly consistent control function projection $\hat{C}(A)$ is available. Furthermore, the $\sqrt{n_2}$-consistency of $\hat{\boldsymbol \alpha}$ requires the uniform convergence rate of $\hat{C} (A)$ to be at least faster than $n_2^{-1/4}$.
In this scenario, estimating $C(A)$ has a negligible impact on the asymptotic variance of $\sqrt{n_2} (\hat{\boldsymbol \alpha} - \boldsymbol \alpha)$. Theorem \ref{thm2} also implies   the asymptotic variance of $\sqrt{n_2} ( \hat{\boldsymbol \alpha} - \boldsymbol \alpha )$ will be the same even if we know the true control function projection $C(A)$.

At the end of this subsection, we provide further discussion about  Assumption \ref{assump:regular1} by specifically considering the case where $m(Z)$ follows a linear model. This choice is particularly useful as it covers all cases where $Z$ is binary. The decision to impose a linear restriction on $m(Z)$ is beneficial because it simplifies the estimation process in Step 1. Without loss of generality, we assume that $\hat{m} ( Z )  =  \hat{\gamma}_0 + \hat{\gamma}_1 Z$, where  $\hat{\gamma}_0$ and $\hat{\gamma}_1$ can be obtained through  linear regression.
However, despite this simplification, there are still numerous challenges that need to be addressed in the following  step.

In Step 2 of our analysis, given the linear model   for $m(Z)$, we can derive the following expression using straightforward algebra:
\begin{equation}
    \label{eq:est-hatcA}
    \hat{C} (A)  = A - \hat{\gamma}_0 - \hat{\gamma}_1 \hat{E} (Z\mid A)
 .
\end{equation}
Estimating  $E(Z\mid A)$ will   be somewhat simpler than directly estimating $C(A)$, given its more direct expression. To derive $\hat{E} (Z\mid A)$ in \eqref{eq:est-hatcA}, we suggest utilizing nonparametric techniques or other machine learning approaches, such as kernel estimation~\citep{gasser1979kernel} and random forests~\citep{breiman2001random}.   Similar  estimation approaches can be employed for further estimating the  $\boldsymbol \alpha$ in Step 3. The asymptotic results for $\hat{ \boldsymbol \alpha}$ are straightforward, as Assumption \ref{assump:regular1} is satisfied by the estimator $\hat{ C}(A)$ constructed in \eqref{eq:est-hatcA}. To summarize:

\begin{corollary}\label{corollary}
 If $\hat{E}(Z \mid A)$ is uniformly consistent and $E(Z\mid A)$ is bounded, then  Assumption \ref{assump:regular1}(i)  holds.
Also, when $\sup_{A\in \mathcal{A}} \lvert \hat{E} (Z\mid A) -E(Z\mid A)  \rvert = o_p ( n_1^{-1/4} )$, Assumption \ref{assump:regular1}(ii)  holds.

\end{corollary}

\section{Simulation Studies}\label{simu}

To evaluate the finite sample performance of the proposed three-step estimator, we firstly generate data according to the following model:
\begin{align*}
  \text{Scenario 1}:  A &= \gamma Z + l U  + \varepsilon,~~  Y =  \alpha A + \beta U + \eta, \\
  \text{Scenario 2}: A &= \gamma Z + l U  + \varepsilon,~~  Y =  \alpha A^2 + \beta U + \eta.
\end{align*}
In Scenario 1, we consider the linear treatment effect setting, while in Scenario 2, we examine the nonlinear treatment effect setting. The goal is to demonstrate the consistency of our proposed estimator in various situations involving different distributional combinations of $U$ and $\varepsilon$, as well as different sample sizes. We have $n_2$ samples for $(A, Y)$ from the primary dataset and another $n_1$ samples for $(Z, A)$ from the auxiliary dataset.
For both scenarios, we consider different combinations of $n_1$ and $n_2$, where $n_1, n_2 \in \{5000, 10000\}$, and we use either a binary or continuous instrument variable $Z$. We first set $ \alpha = \gamma = \beta = 1$ and assume $(U, \varepsilon) \sim \mathrm{N}(0, I_2)$ with $l = 0.5$. In this case, $(U, \varepsilon)$ belongs to an elliptically contoured distribution, satisfying Assumption~\ref{A2} from the previous Example~\ref{exm_2}. Next, we fix $l = 1$ and consider $U, \varepsilon {\sim}  \exp (1)$ or $U(-1,1)$. These distributions also satisfy Assumption~\ref{A2} from Example~\ref{exm_3}. Additional simulation settings, including larger sample sizes or other forms of distribution combinations, are provided in the supplementary material. We also display the simulation results for control function approach with complete dataset at the end of the supplement.

For both scenarios, we employ the estimation method proposed in Section \ref{est} for analysis. Intuitively, the estimation performance will be affected by the choice of the nonparametric estimation basis function, the distributional form of $(Z,A)$, and the sample sizes $n_1$ and $n_2$. Table~\ref{tab:linear} presents the bias, mean squared error (MSE), and $95\% $ coverage probability of our estimator with 500 bootstrap resampling iterations and 500 Monte Carlo runs. Across all scenarios, the bias for our estimator is found to be very small, indicating that it provides unbiased estimates of the causal effect $\boldsymbol \alpha$. Additionally, the coverage probability of our estimator improves as $n_1$ and $n_2$ increase, particularly when $n_2$ is increased. This observation aligns with the convergence performance of $\hat{\boldsymbol \alpha}$ with respect to the sample size $n_2$, as demonstrated in Theorem \ref{thm2}. The  MSE varies under different distributional combinations. For example, when $(U,\varepsilon)$ are normally distributed and $Z$ follows a Bernoulli or uniform distribution, the MSE is relatively larger compared to other cases. This difference in MSE might be attributed to the chosen estimation method for $C(A)$. There is also an interesting observation in  Table~\ref{tab:linear}: Scenario 2, which incorporates nonlinearity, demonstrates superior performance compared to Scenario 1, which adopts a linear treatment model. Overall, as sample size increases, the performance results demonstrate the theoretic results discussed  in Section \ref{est}.

\begin{table}[t]
\centering
\resizebox{0.9\textwidth}{!}{
\begin{threeparttable}
\caption{Bias, mean square error (MSE), and $95\%$ coverage probability ($95\%$  CP) of $\hat{\boldsymbol \alpha}$ based on 500 repetitions. The bias and MSE have been multiplied by $100$.}
\label{tab:linear}
\begin{tabular}{cccccccc}
\toprule
$(n_1,n_2)$ & Metrics& \multicolumn{6}{c}{Scenario 1} \\ \cmidrule(lr){3-8}
& & Setting 1 & Setting 2 & Setting 3 & Setting 4 & Setting 5 & Setting 6 \\\addlinespace[1mm]
& Bias & 0.123 & 0.264 & 0.607 & 0.650 & 0.199 & 0.771 \\
$(5000,5000)$ & MSE & 1.820   & 46.101 & 0.125   & 0.177	 & 1.130  & 48.648 \\
& $95\%$ CP & 95.4 & 97.2 & 94.8 & 93.8 & 96.2 & 98.2 \\\addlinespace[1.5mm]
& Bias & 0.930 & 0.865 & 0.856  & 0.130 & 0.361 & 2.686 \\
$(5000,10000)$ & MSE & 0.971    & 22.769  &	0.084  & 0.110  &  0.622  & 25.968 \\
& $95\%$ CP & 96.0 & 96.6 & 93.4 & 94.4 & 95.4 & 98.0 \\\addlinespace[1.5mm]
& Bias & 1.763  & 3.105	& 0.267  & 0.125  & 1.320  & 3.589 \\
$(10000,5000)$ & MSE & 1.781   & 45.471   & 0.083    & 0.167   & 1.079  & 44.511 \\
& $95\%$ CP & 94.4 & 96.0 & 94.2 & 94.8 & 96.4 & 96.6 \\\addlinespace[1.5mm]
& Bias & 0.497   & 1.482 & 0.511  & 0.145 & 0.002 & 1.589 \\
$(10000,10000)$ & MSE & 0.909    & 25.123  & 0.060      & 0.086   & 0.585   & 24.015 \\
& $95\%$ CP & 96.0 & 94.4 & 95.0 & 95.2 & 95.6 & 96.4 \\ \midrule
$(n_1,n_2)$ & Metrics & \multicolumn{6}{c}{Scenario 2} \\ \cmidrule(lr){3-8}
& & Setting 1 & Setting 2 & Setting 3 & Setting 4 & Setting 5 & Setting 6 \\\addlinespace[1mm]
& Bias & 0.045 & 0.066 & 0.072 & 0.021 & 0.027 & 0.070 \\
$(5000,5000)$ & MSE & 0.013 & 0.008 & 0.002 & 0.002 & 0.012 & 0.007 \\
& $95\%$ CP & 92.4 & 96.0 & 94.2 & 96.6 & 92.2 & 96.0\\\addlinespace[1.5mm]
& Bias & 0.005 & 0.019 & 0.084 & 0.034 & 0.017 & 0.024 \\
$(5000,10000)$ & MSE & 0.007 & 0.004 & 0.001 & 0.001 & 0.006 & 0.004 \\
& $95\%$ CP & 93.2 & 94.8 & 94.4 & 94.8 & 93.4 & 94.8\\\addlinespace[1.5mm]
& Bias & 0.019 & 0.000 & 0.025 & 0.051 & 0.015 & 0.014 \\
$(10000,5000)$ & MSE & 0.012 & 0.009 & 0.001 & 0.002 & 0.012 & 0.008 \\
& $95\%$ CP & 93.0 & 95.0 & 95.8 & 93.0 & 93.2 & 93.6 \\\addlinespace[1.5mm]
& Bias & 0.063 & 0.020 & 0.056 & 0.035 & 0.060 & 0.034 \\
$(10000,10000)$ & MSE & 0.005 & 0.004 & 0.001 & 0.001 & 0.005 & 0.004 \\
& $95\%$ CP & 95.6 & 94.2 & 93.4 & 94.4 & 95.0 & 93.6 \\ \bottomrule
\end{tabular}
\begin{tablenotes}
\item Setting 1 \& 2: $l=1 $, $Z \sim B(0.5)$, $U\sim \exp(1)$;\qquad $l=0.5 $, $Z \sim B(0.5)$,  $U\sim  N(0,1)$;
\item Setting 3 \& 4: $l=1 $, $Z \sim \exp(1)$, $U\sim U(-1,1)$; \quad $l=0.5 $, $Z \sim \exp(1)$,  $U\sim  N(0,1)$;
\item Setting 5 \& 6: $l=1 $, $Z  \sim U(-1,1)$, $U\sim \exp(1)$; \quad$l=0.5 $, $Z \sim U(-1,1)$,  $U\sim  N(0,1)$;
\end{tablenotes}
\end{threeparttable}}
\end{table}

\section{Application}\label{app}
We illustrate the proposed method using two different datasets. The primary dataset $\mathcal{O}_2$ is derived from the National Health and Nutrition Examination Survey (NHANES) program conducted by the Centers for Disease Control and Prevention (CDC) in the United States during 2011-2012. Vitamin D deficiency has been proven to be closely associated with many common diseases, such as diabetes or cancer. In this study, our main focus is on estimating the treatment effect of vitamin D (the treatment $A$) status on BMI (the outcome $Y$), aiming to explore the potential causal relationship between vitamin D deficiency and obesity. However, some crucial confounders, such as gene expressions, which could affect both the vitamin D status and BMI, have not been included in the dataset $\mathcal{O}_2$.  This implies that the treatment effect estimate may suffer from severe bias if these confounders are not considered. After removing missing values  and outliers, our analysis includes {7539} individual samples.

We consider another dataset from the population-based study Monica10 as the auxiliary dataset $\mathcal{O}_1$. The Monica10 study was previously conducted in 1982-1984 and contained examinations of 3785 individuals of Danish origin, recruited from the Danish Central Personal Register. Specifically, this dataset contains five important background variables, including vitamin D status, filaggrin genotype, mortality, age, and follow-up time. The mutations in the filaggrin gene are associated with a higher vitamin D status through an increased UV sensitivity of keratinocytes~\citep{skaaby2013vitamin}. Therefore, we treat the indicator of filaggrin mutation as the instrumental variable ($Z$) since filaggrin plays a crucial role in the skin barrier function but does not seem to affect BMI directly. After removing individuals with missing information  {and outliers},  we finally include 2571 individuals in our analysis ~\citep{martinussen2019instrumental}.

In this section, we apply our method to evaluate the effect of vitamin D status on BMI. The binary instrumental variable $Z$ takes two values, $0$ and $1$, respectively representing the two most common null mutations of the filaggrin gene, including R501X and 2282del4. We employ the estimation method proposed in Section \ref{simu} for our analysis. {{The corresponding assumptions can be empirically verified by the implementation process of our method. For example, Assumption~\ref{A1} can be directly satisfied when the control function projection $C(A)$ is nonlinear in $A$ for linear treatment effect cases.}} The results are shown in Table~\ref{vid_BMI}, where we provide the point estimate of $ \alpha$, {{along with the bootstrap standard errors and 95\% confidence interval calculated using the corresponding $z$-score.  The quantiles and histogram of the corresponding bootstrap results are also provided in the supplementary material. For comparison, we have also included the naive results obtained through the least square in the second row of Table~\ref{vid_BMI}. We note a significant negative effect of vitamin D on BMI using naive estimates. However, vitamin D is widely used in practice, and no studies suggest that vitamin D has significant side effects. Our proposed method}} indicates that vitamin D status has no significant effect on BMI, which is consistent with previous findings reported by~\cite {duan2020effects}. However, possibly due to the lack of important pre-treatment covariates related to distinct timeframes and locations, the primary and auxiliary datasets may not be completely independent and identically distributed. Thus, we suggest practitioners collect more covariates information and further consider Model \eqref{eqn:newmodelforAY} and the methods proposed in Section \ref{relax} for analysis.

\begin{table}[t!]
  \centering
  \setlength{\abovecaptionskip}{0.cm}
%  \caption{\red{Point estimate, the bootstrap 2.5\%, 25\%, 50\%, 75\% and 97.5\% quantile ($Q2.5\%,Q1,Q2,Q3, Q97.5\%$), bootstrap standard error (SE) and 95\% confidence interval (95\% CI using $z$-score) for causal effect of vitamin D status on BMI. }}
    \caption{Point estimate, {{bootstrap}} standard error (SE), and 95\% confidence interval for the causal effect of vitamin D status on BMI. }
  \label{vid_BMI}
  \resizebox{0.908575\columnwidth}{!}{
  \setlength{\tabcolsep}{4mm}
    \begin{tabular}{cccccccc}
    \addlinespace[2mm]
    \toprule
 & Point estimate & SE   & 95\% confidence interval     \\\addlinespace[1mm]
    \cline{2-4}\addlinespace[1mm]
The proposed method  &-0.6386  &14.2409  & (-28.5508, 27.2736) \\\addlinespace[1mm]
Naive method &-0.0418 &0.0034 & (-0.0484, -0.0352)\\
\bottomrule
    \end{tabular} }
\end{table}%

\section{Extension} \label{relax}
In previous sections, we assumed that the selection indicator variable $R \indep (Z,A,Y)$, implying that the corresponding missing mechanism is missing completely at random~\citep{rubin1976inference}. However, even in cases where the selection indicator $R$ depends on  some observed variables, it remains feasible to attain the identification of causal parameter $\boldsymbol \alpha$. We summarize this in the following assumption.
\begin{assumption} \label{A6}
    $R \indep (Z,Y) \mid A$.
\end{assumption}
Assumption \ref{A6} can be viewed as a form of missing at random (MAR) assumption, where the selection indicator $R$ is allowed to depend on treatment variable  $A$.
This assumption covers scenarios in which the two data sources $\mathcal{O}_1$ and $\mathcal{O}_2$ share overlapping samples that provide treatment information, a situation commonly encountered in practice.
%This assumption accommodate the case that the samples of treatment $A$ from $\mathcal{O}_1$ and $\mathcal{O}_2$ overlap, which is also a very common scenario in practice.
Intuitively, since the treatment variable $A$ is allowed to be observed in both datasets, Assumption \ref{A6} can still sufficiently establish identification results similar to Theorem \ref{thm1}.

\begin{theorem}
\label{thm:mar}
    Under Assumptions~\ref{A1},~\ref{A2} and~\ref{A6}, the causal parameter $\boldsymbol \alpha$ can be identified as follows
    $$
\boldsymbol \alpha = \boldsymbol D_{11} E\{ \boldsymbol g(A) E( Y\mid A,R=1 )  \} + \boldsymbol D_{12} E\{ C(A) E(Y\mid A,R=1)  \},
    $$
where the conditional expectation $C(A)$ can be identified through $ C(A) = A - E\{  E(A\mid Z)   \mid A,R=0 \} $, and  the conditional expectation $E(A\mid Z)$ can be identified through $E(A\mid Z=z)=\int a f (a\mid z) d a$ with
$$
f(a\mid z) = \frac{ f (z\mid a, r=0) f (a) } {f (z)}, ~~ f (z) = \int f (z \mid a, r=0) f (a) d a.
$$
\end{theorem}
Based on the identification results presented in Theorem \ref{thm:mar}, we can develop an estimation procedure for causal parameter $\boldsymbol \alpha$ and provide its asymptotic properties using a similar approach as described in Section \ref{est}. For the sake of simplicity, we omit the detailed methodology in this section.

\section{Discussion} \label{dis}
In this article, we  consider identifying and estimating the causal effect using an instrumental variable from auxiliary dataset. Existing researches often rely on joint observations of the instrumental variable and outcome, such as two-sample instrumental estimators, to identify treatment effects. However, this poses challenges when the instrumental variable is not available in the primary dataset.
To address this issue, we propose a novel identification strategy  from the control function perspective.   We further consider the estimation and asymptotic theory for the proposed estimator.

There are several potential extensions that can be explored in future researches. Firstly, if additional information, such as proxy variables or multiple candidate instrumental variables, becomes available, the identification conditions may be further relaxed or extended to allow for nonparametric identification~\citep{kang2016instrumental,miao2018identifying}. Secondly, the scenario with binary treatment variables or non-continuous outcome variables is pretty common in practice, and it would be interesting to extend our model to accommodate such cases \citep{wang2018bounded}. The difficulty is about how to propose a similar condition like the proportional conditional expectation equality under linear additive model. Finally, investigating causal inference in settings with multiple or high-dimensional treatments holds practical and theoretical importance.  These are all left for future work because they are beyond the scope of this paper.

 \section*{Acknowledgments}
We sincerely thank the editor, associate editor, and reviewers for their insightful and helpful comments, which have significantly improved our paper.  Kang Shuai and Yangbo He are supported by the National Key R\&D Program of China (2022ZD0160300). Wei Li is supported by the Beijing Natural Science Foundation (1232008), the National Natural Science Foundation of China (12101607, 12071015),  the National Key R\&D Program of China (2022YFA1008100), and the MOE Project of Key Research Institute of Humanities and Social Sciences (22JJD910001).  Shanshan Luo is supported by the Disciplinary funding of Beijing Technology and Business University and the Research Foundation for Youth Scholars of Beijing Technology and Business University.

\section*{Supplementary Material}
\hspace{0.6cm}
The supplementary material available online includes  %a heuristic discussion,
 additional technical proofs and simulation results.

\begin{spacing}{1.25}

\bibhang=1.7pc
\bibsep=2pt
\fontsize{9}{14pt plus.8pt minus .6pt}\selectfont
\renewcommand\bibname{\large \bf References}
%\begin{thebibliography}{11}
\expandafter\ifx\csname
natexlab\endcsname\relax\def\natexlab#1{#1}\fi
\expandafter\ifx\csname url\endcsname\relax
  \def\url#1{\texttt{#1}}\fi
\expandafter\ifx\csname urlprefix\endcsname\relax\def\urlprefix{URL}\fi

%% use bibfile
  \bibliographystyle{plainnat}      % Chicago style, author-year citations
  \bibliography{ref.bib}   % name your BibTeX data base

@incollection{heckman1976common,
  title={The common structure of statistical models of truncation, sample selection and limited dependent variables and a simple estimator for such models},
  author={Heckman, James J},
  booktitle={Annals of Economic and Social Measurement, volume 5},
  pages={475--492},
  year={1976},
  publisher={NBER}
}

@article{goldberger1972structural,
  title={Structural equation methods in the social sciences},
  author={Goldberger, Arthur S},
  journal={Econometrica: Journal of the Econometric Society},
  volume={40},
  pages={979--1001},
  year={1972},
  publisher={JSTOR}
}

@book{wooldridge2010econometric,
  title={Econometric Analysis of Cross Section and Panel Data},
  author={Wooldridge, Jeffrey M.},
  year={2010},
  publisher={MIT Press}
}

@article{hansen1982large,
  title={Large Sample Properties of Generalized Method of Moments Estimators},
  author={Hansen, Lars Peter},
  journal={Econometrica},
  volume={50},
  pages={1029--1054},
  year={1982}
}

@article{chernozhukov2018double,
  title={Double/debiased machine learning for treatment and structural parameters: Double/debiased machine learning},
  author={Chernozhukov, Victor and Chetverikov, Denis and Demirer, Mert and Duflo, Esther and Hansen, Christian and Newey, Whitney and Robins, James},
  journal={The Econometrics Journal},
  volume={21},
  year={2018},
  pages={C1 -- C68},
  publisher={Oxford University Press}
}

@unpublished{petrin2011revisiting,
  note = {NBER Working Paper Series},
  title={REVISITING INSTRUMENTAL VARIABLES AND THE CLASSIC CONTROL FUNCTION APPROACH, WITH IMPLICATIONS FOR PARAMETRIC AND NON-PARAMETRIC REGRESSIONS Kyoo il Kim},
  author={Petrin, Amil},
  year={2011}
}

@article{guo2016control,
  title={Control function instrumental variable estimation of nonlinear causal effect models},
  author={Guo, Zijian and Small, Dylan S},
  journal={The Journal of Machine Learning Research},
  volume={17},
  pages={3448--3482},
  year={2016},
  publisher={JMLR. org}
}

@article{rubin1980randomization,
  title={Randomization analysis of experimental data: The {F}isher randomization test comment},
  author={Rubin, Donald B},
  journal={Journal of the American Statistical Association},
  volume={75},
  pages={591--593},
  year={1980},
  publisher={JSTOR}
}

@book{fang2018symmetric,
  title={Symmetric multivariate and related distributions},
  author={Fang, Kai-Tai and Kotz, Samuel and Ng, Kai Wang},
  year={2018},
  publisher={Chapman and Hall/CRC}
}

@article{owen1983class,
  title={On the class of elliptical distributions and their applications to the theory of portfolio choice},
  author={Owen, Joel and Rabinovitch, Ramon},
  journal={The Journal of Finance},
  volume={38},
  pages={745--752},
  year={1983},
  publisher={Wiley Online Library}
}

@article{hahn2018nonparametric,
  title={Nonparametric two-step sieve M estimation and inference},
  author={Hahn, Jinyong and Liao, Zhipeng and Ridder, Geert},
  journal={Econometric Theory},
  volume={34},
  pages={1281--1324},
  year={2018},
  publisher={Cambridge University Press}
}

@article{angrist1996identification,
  title={Identification of causal effects using instrumental variables},
  author={Angrist, Joshua D and Imbens, Guido W and Rubin, Donald B},
  journal={Journal of the American Statistical Association},
  volume={91},
  pages={444--455},
  year={1996},
  publisher={Taylor \& Francis}
}

@incollection{imbens2007control,
  title={Control function and related methods},
  author={Imbens, Guido and Wooldridge, Jeffrey},
  booktitle={What’s new in Econometrics},
  year={2007},
  publisher={NBER SummerInstitute}
}

@article{li2020causal,
  title={Causal inference for nonlinear outcome models with possibly invalid instrumental variables},
  author={Li, Sai and Guo, Zijian},
  journal={arXiv preprint arXiv:2010.09922},
  year={2020}
}

@article{rivers1988limited,
  title={Limited information estimators and exogeneity tests for simultaneous probit models},
  author={Rivers, Douglas and Vuong, Quang H},
  journal={Journal of Econometrics},
  volume={39},
  pages={347--366},
  year={1988},
  publisher={Elsevier}
}

@article{cai2011two,
  title={Two-stage instrumental variable methods for estimating the causal odds ratio: analysis of bias},
  author={Cai, Bing and Small, Dylan S and Have, Thomas R Ten},
  journal={Statistics in Medicine},
  volume={30},
  pages={1809--1824},
  year={2011},
  publisher={Wiley Online Library}
}

@article{ogburn2015doubly,
  title={Doubly robust estimation of the local average treatment effect curve},
  author={Ogburn, Elizabeth L and Rotnitzky, Andrea and Robins, James M},
  journal={Journal of the Royal Statistical Society: Series B (Statistical Methodology)},
  volume={77},
  pages={373--396},
  year={2015},
  publisher={Wiley Online Library}
}

@article{wang2018bounded,
  title={Bounded, efficient and multiply robust estimation of average treatment effects using instrumental variables},
  author={Wang, Linbo and Tchetgen Tchetgen, Eric},
  journal={Journal of the Royal Statistical Society: Series B (Statistical Methodology)},
  volume={80},
  pages={531--550},
  year={2018},
  publisher={Wiley Online Library}
}

@article{shuai2024identification,
  title={Identification and Estimation of Causal Effects Using non-Gaussianity and Auxiliary Covariates},
  author={Shuai, Kang and Luo, Shanshan and Zhang, Yue and Xie, Feng and He, Yangbo},
  journal={to appear in Statistica Sinica},
  year={2024}
}

@article{skaaby2013vitamin,
  title={Vitamin D status, filaggrin genotype, and cardiovascular risk factors: a Mendelian randomization approach},
  author={Skaaby, Tea and Husemoen, Lise Lotte Nystrup and Martinussen, Torben and Thyssen, Jacob P and Melgaard, Michael and Thuesen, Betina Heinsb{\ae}k and Pisinger, Charlotta and J{\o}rgensen, Torben and Johansen, Jeanne D and Menn{\'e}, Torkil and others},
  journal={PloS one},
  volume={8},
  pages={e57647},
  year={2013},
  publisher={Public Library of Science San Francisco, USA}
}

@article{arellano1992female,
  title={Female labour supply and on-the-job search: an empirical model estimated using complementary data sets},
  author={Arellano, Manuel and Meghir, Costas},
  journal={The Review of Economic Studies},
  volume={59},
  pages={537--559},
  year={1992},
  publisher={Wiley-Blackwell}
}

@article{gamazon2015gene,
  title={A gene-based association method for mapping traits using reference transcriptome data},
  author={Gamazon, Eric R and Wheeler, Heather E and Shah, Kaanan P and Mozaffari, Sahar V and Aquino-Michaels, Keston and Carroll, Robert J and Eyler, Anne E and Denny, Joshua C and GTEx Consortium and Nicolae, Dan L and others},
  journal={Nature Genetics},
  volume={47},
  pages={1091--1098},
  year={2015},
  publisher={Nature Publishing Group US New York}
}

@article{zhao2019two,
  title={Two-sample instrumental variable analyses using heterogeneous samples},
  author={Zhao, Qingyuan and Wang, Jingshu and Spiller, Wes and Bowden, Jack and Small, Dylan S},
  journal={Statistical Science},
  volume={34},
  pages={317--333},
  year={2019},
  publisher={JSTOR}
}

@book{kotz2004continuous,
  title={Continuous multivariate distributions, Volume 1: Models and applications},
  author={Kotz, Samuel and Balakrishnan, Narayanaswamy and Johnson, Norman L},
  volume={1},
  year={2004},
  publisher={John Wiley \& Sons}
}

@inproceedings{gasser1979kernel,
  title={Kernel estimation of regression functions},
  author={Gasser, Theo and M{\"u}ller, Hans-Georg},
  booktitle={Smoothing Techniques for Curve Estimation: Proceedings of a Workshop Held in Heidelberg},
 pages={23--68},
  year={1979},
  organization={Springer}
}

@article{breiman2001random,
  title={Random forests},
  author={Breiman, Leo},
  journal={Machine Learning},
  volume={45},
  pages={5--32},
  year={2001},
  publisher={Springer}
}

@article{rubin1976inference,
  title={Inference and missing data},
  author={Rubin, Donald B},
  journal={Biometrika},
  volume={63},
  pages={581--592},
  year={1976},
  publisher={Oxford University Press}
}

@article{miao2018identifying,
  title={Identifying causal effects with proxy variables of an unmeasured confounder},
  author={Miao, Wang and Geng, Zhi and Tchetgen Tchetgen, Eric J},
  journal={Biometrika},
  volume={105},
  pages={987--993},
  year={2018},
  publisher={Oxford University Press}
}

@article{kang2016instrumental,
  title={Instrumental variables estimation with some invalid instruments and its application to Mendelian randomization},
  author={Kang, Hyunseung and Zhang, Anru and Cai, T Tony and Small, Dylan S},
  journal={Journal of the American Statistical Association},
  volume={111},
  pages={132--144},
  year={2016},
  publisher={Taylor \& Francis} 
}

@article{petrin2010control,
  title={A control function approach to endogeneity in consumer choice models},
  author={Petrin, Amil and Train, Kenneth},
  journal={Journal of Marketing Research},
  volume={47},
  pages={3--13},
  year={2010},
  publisher={SAGE Publications Sage CA: Los Angeles, CA}
}

@article{wooldridge2015control,
  title={Control function methods in applied econometrics},
  author={Wooldridge, Jeffrey M},
  journal={Journal of Human Resources},
  volume={50},
  pages={420--445},
  year={2015},
  publisher={University of Wisconsin Press}
}

@book{angrist2013advances,
  title={Advances in Economics and Econometrics},
  author={Angrist, Joshua and Fernandez-Val, Ivan and Acemoglu, Daron and Arellano, Manuel and Eddie, D},
  year={2013},
  publisher={Cambridge University Press}
}

@article{martinussen2019instrumental,
  title={Instrumental variables estimation under a structural Cox model},
  author={Martinussen, Torben and N{\o}rbo S{\o}rensen, Ditte and Vansteelandt, Stijn},
  journal={Biostatistics},
  volume={20},
  pages={65--79},
  year={2019},
  publisher={Oxford University Press}
}

@article{duan2020effects,
  title={Effects of vitamin D supplementation on general and central obesity: results from 20 randomized controlled trials involving apparently healthy populations},
  author={Duan, Leizhen and Han, Ling and Liu, Qin and Zhao, Yili and Wang, Lei and Wang, Yan},
  journal={Annals of Nutrition and Metabolism},
  volume={76},
  pages={153--164},
  year={2020},
  publisher={S. Karger AG Basel, Switzerland}
}

@article{angrist1992effect,
  title={The effect of age at school entry on educational attainment: an application of instrumental variables with moments from two samples},
  author={Angrist, Joshua D and Krueger, Alan B},
  journal={Journal of the American Statistical Association},
  volume={87},
  pages={328--336},
  year={1992},
  publisher={Taylor \& Francis}
}

@article{inoue2010two,
  title={Two-sample instrumental variables estimators},
  author={Inoue, Atsushi and Solon, Gary},
  journal={The Review of Economics and Statistics},
  volume={92},
  pages={557--561},
  year={2010},
  publisher={The MIT Press}
}

@article{zhao2020statistical,
  title={Statistical inference in two-sample summary-data Mendelian randomization using robust adjusted profile score},
  author={Zhao, Qingyuan and Wang, Jingshu and Hemani, Gibran and Bowden, Jack and Small, Dylan S},
  journal={Annals of Statistics},
  year={2020},
  volume={48},
  pages={1742--1769}
}

@article{sun2022semiparametric,
  title={On semiparametric instrumental variable estimation of average treatment effects through data fusion},
  author={Sun, BaoLuo and Miao, Wang},
  journal={Statistica Sinica},
  volume={32},
  pages={569--590},
  year={2022}
}

\end{spacing}

%%  Another method

%%%%%%%%%%%%%%%%%%%%%%%%%%%%%%%%%%%%%%%%%%%%%%%%%%%%%%%%%%%%%%%%%%%%%%%%%%%%%%%%%%%%%%%%%%%%%%%%%%%%%%%%%%%%%%%%%%%%%%%%%%%%

%%%%%%%%%%%%%%%%%%%%%%%%%%%%%%%%%%%%%%%%%%%%%%%%%%%%%%%%%%%%%%%%%%%%%%%%%%%%%%%%%%%%%%%%%%%%%%%%%%%%%%%%%%%%%%%%%%%%%%%%%%%%
%\vskip .65cm
%\noindent
%$^{a}$Beijing International Center for Mathematical Research, Peking University,  Beijing 100871, China
%
%\vskip 2pt
%\noindent
%E-mail: (first author email)
%\vskip 2pt
%
%\noindent
%second author affiliation
%\vskip 2pt
%\noindent
%E-mail: (second author email)

% \vskip .3cm
%\centerline{(Received ???? 20??; accepted ???? 20??)}\par
\end{document}